\documentclass{svproc}
\usepackage{url}
\usepackage{amsmath}
\usepackage{amssymb}
\usepackage{graphicx}
\usepackage{float}
\usepackage{array}
\usepackage{subcaption}

\begin{document}
\mainmatter              
\title{Effective Lepton Flavor Violating couplings at Muon Collider}
\titlerunning{LFV at muon vollider}  
%
\author{Sukanta Dutta\inst{1, 2}  \and
Purnath Unnikrishnan  $^{\star ,}$\inst{3} \and Yashasvi  
\thanks{ Speakers  at the DAE-BRNS HEP Symposium, 19-23, December,2024, BHU, India.}$^,$  \inst{4}}
\authorrunning{Sukanta Dutta et al.} 

\institute{SGTB Khalsa College, University of Delhi, Delhi, India.
\and
Delhi School of Analytics, Institution of Eminence, University of Delhi, Delhi.\\
\email{sukanta.dutta@gmail.com}
\and 
Department of Physics and Astrophysics, University of Delhi, Delhi, India.\\
\email{purnathuk@gmail.com}
\and
Department of Physics and Astrophysics, University of Delhi, Delhi, India.\\
\email{yamansangwan16@gmail.com}}
\maketitle              
\begin{abstract}
We estimate the sensitivity of Wilson coefficients of the lepton flavor-violating dimension-six operators at the proposed $\mu^+\mu^-$ collider. We compute the signal significance at $\sqrt{s}$ = 3 and 10 TeV, respectively, with an integrated luminosity of 1 and 10 ab$^{-1}$ corresponding to unpolarized and polarized initial muon beams. Using the optimal observable method for the kinematic distributions, we study the measurement errors of the effective couplings at the 1-sigma level.
\end{abstract}
\section{The effective Interaction Lagrangian}
The Standard Model (SM) of particle physics is highly successful but challenged to explain the observed baryon asymmetry, neutrino oscillation, absence of dark matter candidates, etc. Although lepton flavor violation (LFV) is absent in the SM, a small contribution to the LFV couplings may arise either in the beyond SM scenarios or through the low energy effective interactions of SM particles.  
\par The signature for LFV has been studied at prospective $e^+\,e^-$  collider \cite{Jahedi:2024kvi}. In this article, we probe the potential of detecting the signatures of the LFV  induced by the dimension six effective four fermionic operators at the proposed muon collider \cite{MuonCollider:2022xlm}. The interaction Lagrangian is given as
\begin{eqnarray} 
& \mathcal{L}_{\text{LFV}} = \left( \frac{C_{\text{LR}}^S}{\Lambda^2}\right)^{ijkl} \ (\bar{\ell_i} P_L \ell_j)\ (\bar{\ell}_k P_R \ell_l) 
+\left(\frac{C_{\text{RL}}^S}{\Lambda^2}\right)^{ijkl} \ (\bar{\ell_i} P_R \ell_j)\ (\bar{\ell}_k P_L \ell_l) \nonumber\\ 
&+  \left( \frac{C_{\text{LL}}^V}{\Lambda^2}\right)^{ijkl}  (\bar{\ell_i} \gamma^\sigma P_L \ell_j)\ (\bar{\ell}_k \gamma_\sigma P_L \ell_l)
+\left(\frac{C_{\text{LR}}^V}{\Lambda^2}\right)^{ijkl}  (\bar{\ell_i} \gamma^\sigma P_L \ell_j)\ (\bar{\ell}_k \gamma_\sigma P_R \ell_l)\nonumber \\
&+  \left(\frac{C_{\text{RL}}^V}{\Lambda^2}\right)^{ijkl} (\bar{\ell_i} \gamma^\sigma P_R \ell_j)\ (\bar{\ell}_k \gamma_\sigma P_L \ell_l) 
+ \left(\frac{C_{\text{RR}}^S}{\Lambda^2}\right)^{ijkl} (\bar{\ell_i} \gamma^\sigma P_R \ell_j)\ (\bar{\ell}_k \gamma_\sigma P_R \ell_l)
\end{eqnarray}
where $i,\,j,\,k,\, l$ are the flavor index of the leptons and subscripts $L$ ($R$) corresponds to the lepton's left (right) chiral current. The muon collider, however, restricts the first two flavor indices to be of the same generation and henceforth we will tag the effective couplings with the last two flavor indices only. We take all these couplings to be real and process them with no final state muon.  Using the Fierz identity, we  further reduce the six effective  couplings into three linearly independent Wilson coefficients: 
\begin{eqnarray}
\left(\frac{C_{LR}}{\Lambda^2}\right) = 
\left(\frac{C_{\text{LR}}^V}{\Lambda^2}\right)^{e\,\tau}= \left(\frac{C_{\text{RL}}^V}{\Lambda^2}\right)^{e\,\tau} && = \,-\frac{1}{2} \left(\frac{C_{\text{LR}}^S}{\Lambda^2}\right)^{ e\,\tau}= \,-\frac{1}{2} \left(\frac{C_{\text{RL}}^S}{\Lambda^2}\right)^{e\,\tau},
\nonumber\\
\left(\frac{C_{LL}}{\Lambda^2}\right) = \frac{1}{4}\left(\frac{C_{\text{LL}}^V}{\Lambda^2}\right)^{e\,\tau} &&\text{and}\, \left(\frac{C_{RR}}{\Lambda^2}\right) = \frac{1}{4}\left(\frac{C_{\text{RR}}^V}{\Lambda^2}\right)^{e\,\tau}.
\end{eqnarray}
\vspace{-1cm}
\section{Collider simulation and Significance}
We investigate the LFV processes induced by the dimension six operators
\begin{eqnarray}
 \mu^+ \mu^- \rightarrow e^{\pm} \tau^{\mp}   
\end{eqnarray}
The  prominent background processes taken into account for the analysis are  
\begin{eqnarray} \mu^+ \mu^- \rightarrow  \tau^{+} \tau^{-},\,\,  \mu^+ \mu^- \rightarrow  W^{+} W^{-},\,\, \text{and} \,\,\mu^+ \mu^- \rightarrow \nu \,\bar{\nu} \,Z
\end{eqnarray}
In the above processes, the tau decays in the hadronic channel. We have implemented the effective interaction Lgarnagian in  \textsc{FeynRules} \cite{Alloul:2013bka} and fed the Feynman rules to the event generator \textsc{MadGraph} \cite{Alwall:2011uj}. The generated events for the backgrounds and signal processes are passed to   \textsc{Pythia8} \cite{Sjostrand:2014zea} for parton showering and then to \textsc{Delphes3} \cite{deFavereau:2013fsa} for detector simulation.
\par
\begin{figure}[H]
    \centering
    \begin{subfigure}[t]{0.55\textwidth}
       
        \renewcommand{\arraystretch}{2}
        \vspace{-5cm}
        \resizebox{\linewidth}{!}{%
                \begin{tabular}{|c|c|c|c|}\hline
            Couplings (GeV$^{-2}$)& $P_{\mu^-}=0\%$& $P_{\mu^-}=-80\%$& $P_{\mu^-}=80\%$\\
            \hline
            $ C_{LL}/\Lambda^2=1 \times 10^{-9}$& 0.74& 1.3& 0.15\\
            \hline
            $ C_{LR}/\Lambda^2=1 \times 10^{-9}$& 0.37& 0.37& 0.37\\
            \hline
            $ C_{RR}/\Lambda^2=1 \times 10^{-9}$& 0.74& 0.15& 1.3\\
            \hline
            Background & 2554&4588&527\\ \hline
        \end{tabular}}
        \centering
        \vspace{0.85cm}
  \subcaption{\em{Cross section of LFV and combined background processes  in $fb$ }}
  \label{compareXsec}
    \end{subfigure}
        \hfill
    \begin{subfigure}{0.4\textwidth}
        \centering
        \includegraphics[width=\linewidth]{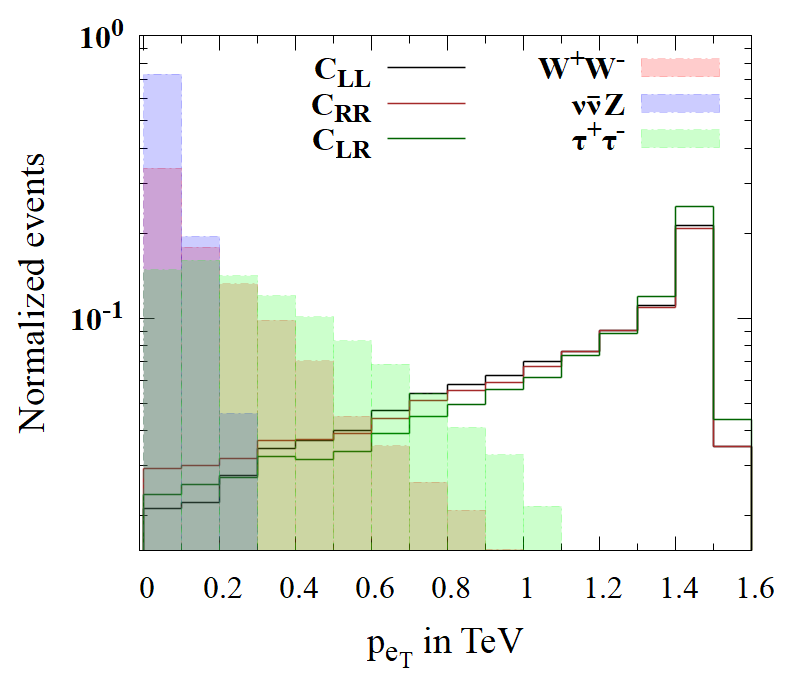}
        \subcaption{$p_T$ distribution of electron in GeV}
        \label{dist}
    \end{subfigure}
    \caption{\em{Results from Collider simulation at $\sqrt{s}$ = 3 TeV with $\mathcal{L}_\text{int} = 1\, \text{ab}^{-1}$ }}
\end{figure}

 Translating the existing upper bounds on flavor violating tau decay $\mathcal{B}(\tau \to \mu\mu e) \le 2.7 \times 10^{-8}$ \cite{Hayasaka:2010np} on the effective couplings, we display in table \ref{compareXsec} the comparative estimation of the cross-sections using unpolarized ($P_{\mu^-}=0\%$) and $\pm 80\%$ polarized muon beams at c.m. energy of 3 TeV for the backgrounds and LFV processes respectively. 
\vspace{-0.6cm}
\hspace{-0.5cm}Examining the $p_T$ distributions of the outgoing lepton from the signal and background processes, as depicted in figure \ref{dist}, we impose the following cuts to minimize the loss of signal while effectively eliminating most of the background contributions: (a) $\text{(veto cuts)}: N_e = 1,\, N_{\tau_h} = 1,\, N_\mu=0$ and (b) $p_{e_T}>1\, TeV$.
where $N_{e}$, $N_{\mu}$, $N_{\tau_h}$ denote the number of electrons, muons and hadronically decayed taus in the final state. 
\par
\begin{figure}[H]
\centering
        \begin{subfigure}{0.4\textwidth}
        \includegraphics[width=\linewidth]{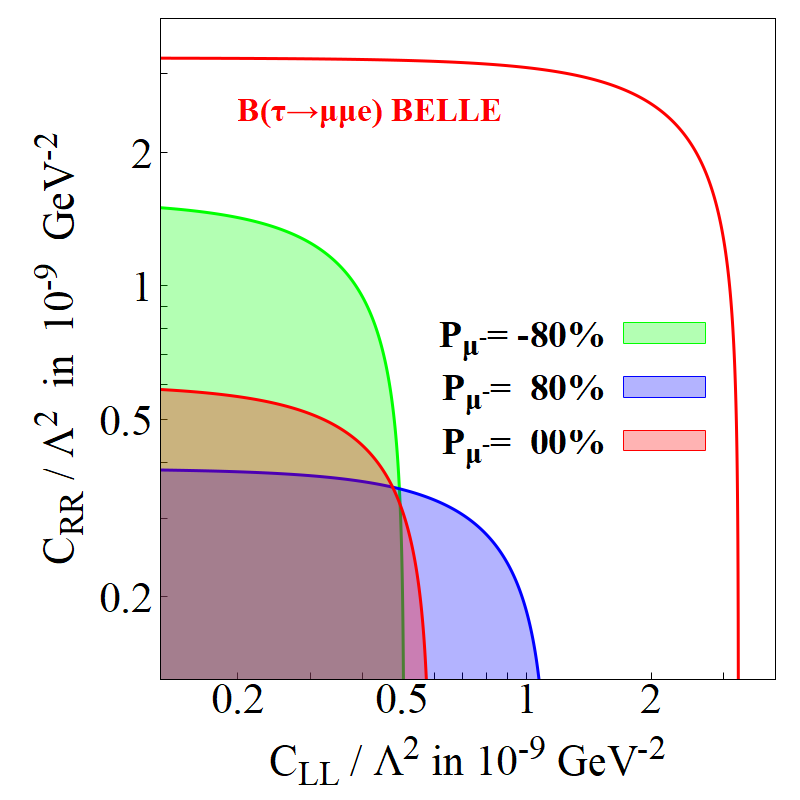}  
        \label{sigCllvsCee}
        \end{subfigure}
        \hspace{0.5cm}
        \begin{subfigure}{0.4\textwidth}
        \includegraphics[width=\linewidth]{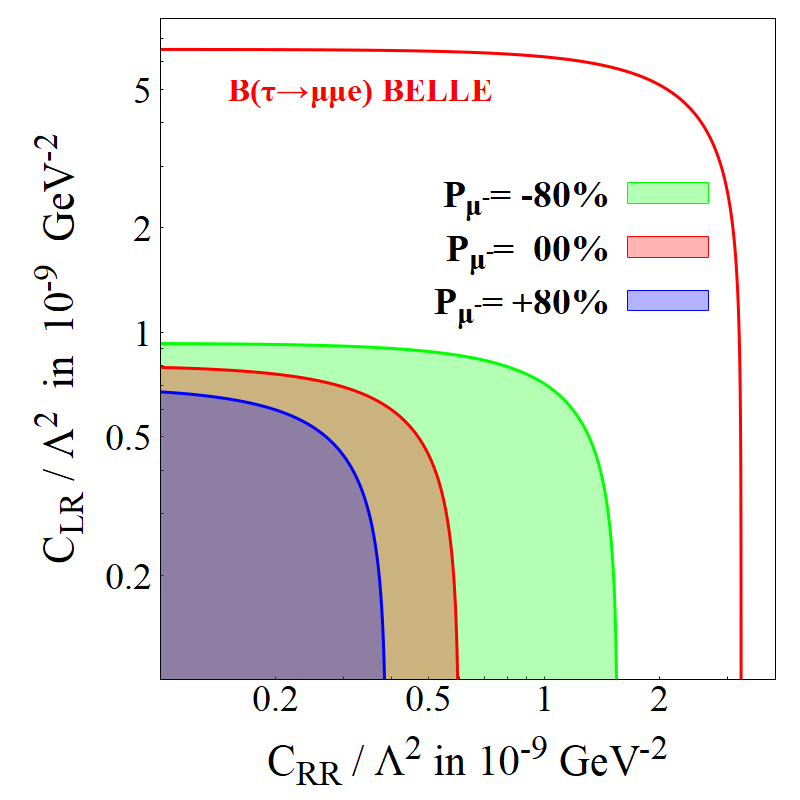}
        \label{sigCeevsCle}
        \end{subfigure}
        \vspace{-0.5cm}
        \caption{\em{$5\,\sigma$ significance contours are depicted for different initial muon beam polarizations in a plane of Wilson coefficients. The red contour corresponds to the upper limit on the  $\mathcal{B}(\tau \to \mu\mu e) \le 2.7 \times 10^{-8}$ from the BELLE experiment \cite{Hayasaka:2010np}.}}
        \label{signif}
\end{figure}

 In figure \ref{signif} we display the 5 $\sigma$ level significance $ \mathcal{S} = \frac{N_S}{\sqrt{N_B}}$ contour plots in the plane defined by the Wilson coefficients for $\sqrt{s}$ = 3 TeV at ${\cal L}_{\rm int}$  of 1 ab$^{-1}$.
\vspace{-0.5cm}
\vspace{-1cm}
\section{Optimal Variable Analysis and Observations}
\begin{figure}[h!]
\centering
        \begin{subfigure}{0.35\textwidth}        \includegraphics[width=\linewidth]{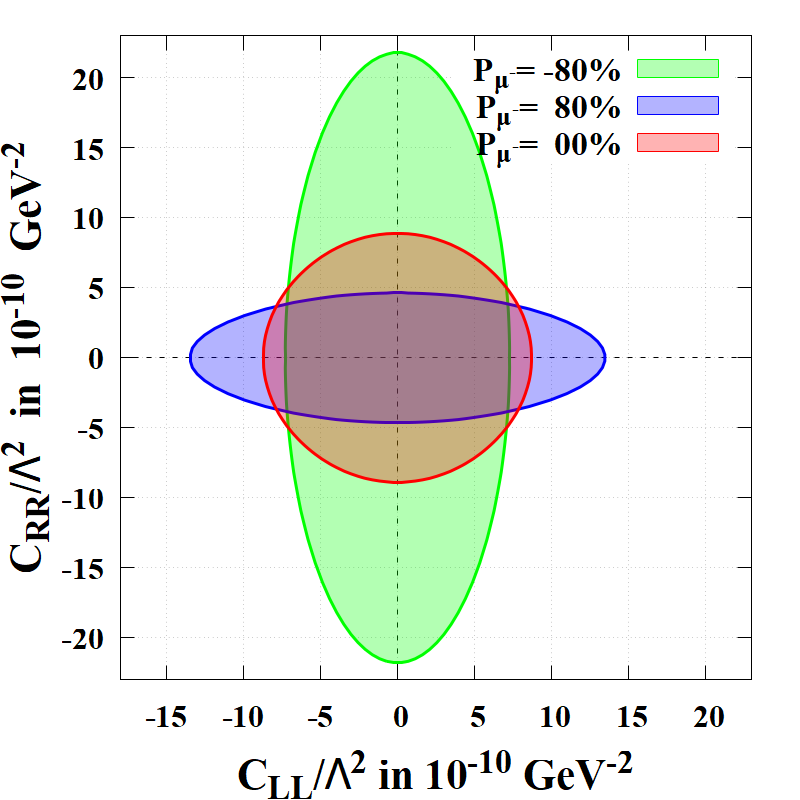}  
        \label{3chi}
        \end{subfigure}
        \hspace{0.5cm}
        \begin{subfigure}{0.35\textwidth}
        \includegraphics[width=\linewidth]{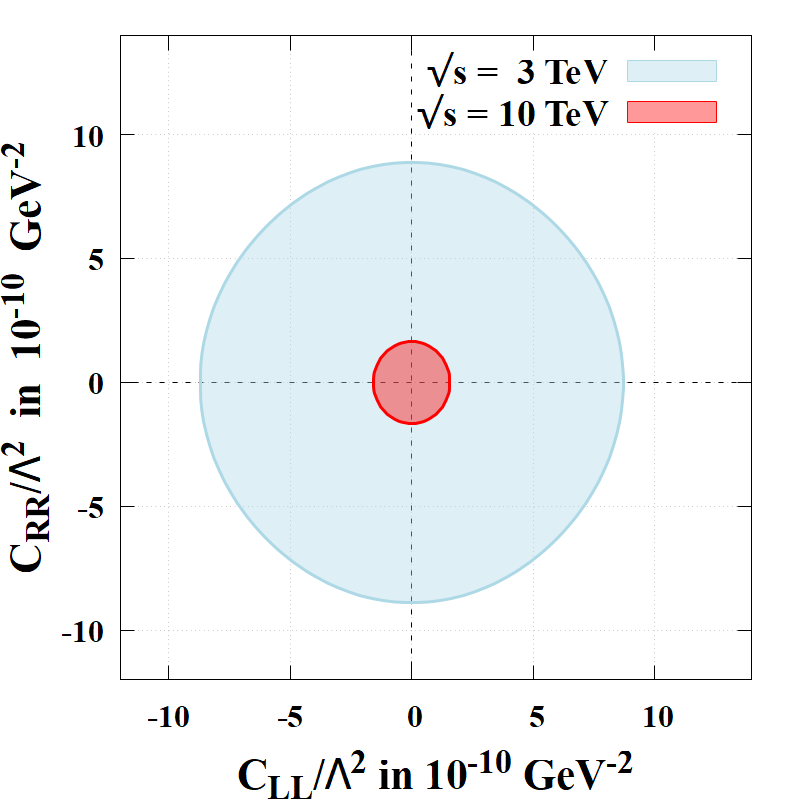} 
        \label{3vs10}
        \end{subfigure}
        \vspace{-0.5em}
        \caption{\em{$\Delta\chi^2=2.3$ contours for $\sqrt{s}=3$ TeV at $\mathcal{L}_{int}=1 $ ab$^{-1}$ at different polarizations(left) and a comparison between 3 TeV at 1 ab $^{-1}$ and 10 TeV at 10 ab$^{-1}$ (right)}  }
        \label{chi2}
    \end{figure}

In the optimal observables method, we make full use of the shape profile of the differential distribution to constrain $c_i$ \cite{Gunion:1996vv}. 
If the $\mu^-$ and $\mu^+$ beam polarizations are $P$ and $\bar{P}$
($|P|,|\bar{P}| \le 1$) respectively, the differential distribution of events can
be expressed as 
\begin{eqnarray}
\frac{d^2N(P,\bar{P})}{d(\cos\theta) d(p_T)} &=& \frac{d^2N_{\rm Bkg.}(P,\bar{P})}{d(\cos\theta) d(p_T)}  \\
&+&\displaystyle\sum_{i} c_i^2 \frac{d^2N_{c_i, c_i}(P,\bar{P})}{d(\cos\theta) d(p_T)}  + 2 \displaystyle\sum_{i} \displaystyle\sum_{j>i} c_i c_j \frac{d^2N_{c_i, c_j}(P,\bar{P})}{d(\cos\theta) d(p_T)}.\nonumber
\end{eqnarray}
For a given integrated Luminosity ${\cal L}$, the inverse of the covariance \( V_{ij} \) matrix   is obtained by differentiating the distribution {\it w.r.t.} \( c_i \) and \( c_j \) respectively.   The expression for the change in \(\chi^2\) for any two effective couplings  \(c_i\) and \(c_j\) is:
\vspace{-0.2em}
\begin{eqnarray} \Delta \chi^2(c_i, c_j) =  \sum_{ij} (c_i - c_i^0)V_{ij}(c_j - c_j^0),\end{eqnarray}
\vspace{-0.2em}
where  \(c_i^0\) and \(c_j^0\) are the best-fit values. The contours with \( \Delta \chi^2 = 2.3 \) represent the one-sigma confidence region for the parameters in figure \ref{chi2} (left) for $\sqrt{s}$ = 3 TeV and figure \ref{chi2} (right) shows a comparison between unpolarized 3 TeV at 1 ab$^{-1}$  and 10 TeV at 
 10 ab$^{-1}$. 
 \vspace{-0.5cm}
    \vspace{-1.1cm}
\subsection{Analysis Summary}
Our analysis shows that the effective LFV  vertices at the muon collider can be probed to very high accuracy at $\sqrt{s}$ of 3 TeV and ${\cal L}_{\rm int} \sim$ 1 ab$^{-1}$ in comparison to existing limits from the LHC, electroweak physics, and  $B$ meson decays. figure \ref{chi2} (right)  shows that we can obtain a comparatively one order or more stringent upper limit on  Wilson's coefficients with $\sqrt{s}$ of 10 TeV and ${\cal L}_{\rm int} \sim$ 10 ab$^{-1}$. Polarization of the initial muon beam is anticipated to play a significant role in enhancing the sensitivity to these effective couplings.\\
{\bf Acknowledgment:} The authors thank Debajyoti Choudhury for the discussions. The authors acknowledge the partial financial support from the ANRF project CRG/2023/008234.
\vspace{-0.5em}

\end{document}